\documentclass[letter,11pt]{article}
\pdfoutput=1

\usepackage{jheppub-mod} 

\usepackage[T1]{fontenc} 
\usepackage{amsthm}

\usepackage{listings}
\usepackage{mathrsfs}

\usepackage{xcolor} 

\hypersetup{pdftitle={Holography and Conditional and Multipartite Entanglements of Purification}, pdfauthor={Ning Bao, Illan F. Halpern}, citecolor=blue,linkcolor=blue,urlcolor=blue}

\newcommand{\be}{\begin{equation}}
\newcommand{\ee}{\end{equation}}

\DeclareMathOperator{\Ar}{Area}

\usepackage{color}
\definecolor{purple}{rgb}{0.5,0,0.5}

\title{Conditional and Multipartite Entanglements of Purification and Holography}

\author{Ning Bao}
\author{and Illan F. Halpern}
\affiliation{Berkeley Center for Theoretical Physics,\\University of California, Berkeley, CA 94720, USA}

\emailAdd{ningbao75@gmail.com}
\emailAdd{illan@berkeley.edu}

\abstract{In this work we generalize the entanglement of purification and its conjectured holographic dual to conditional and multipartite versions of the same, where the optimization defining the entanglement of purification is now optimized in either a constrained way or over multiple parties. We separately derive new constraints on both the conditional entanglement of purification and its conjectured holographic dual object that match, further reinforcing the likelihood of this conjecture. We also  show that the multipartite objects we define, despite obeying several of the same inequalities, are not holographic duals of each other. Further, we find inequalities that are true only for the bulk objects, and thus could provide additional consistency checks for states dual to (semi)-classical bulk geometries.}

\begin{document} 
\maketitle
\flushbottom

\section{Introduction}
There has been much progress made in recent years at the intersection of quantum information and quantum gravity. One particular area of impact is the study of entanglement entropy in the context of holography pioneered by \cite{Ryu:2006bv}. In this context, it was discovered that entanglement entropies for states in holographic theories dual to a classical bulk geometry obeyed additional inequalities beyond those obeyed by all quantum states \cite{Hayden:2011ag, EntCone} and that such states are a small fraction of the set of all quantum states in the entropy space measure \cite{bao2017holographic}. These inequalities, though not true for all quantum systems, therefore serve as a useful discriminator for which quantum states, even in theories known to possess a holographic duality, can be dual to (semi)-classical spacetimes.

It is therefore a natural question to ask whether other entanglement measures are also dual to objects in holography. Recently, it has been conjectured by \cite{TakUme, Nguyen} that the entanglement of purification ($E_p$) \cite{Terhal} is dual to an object called the entanglement wedge cross-section ($E_W$). This conjecture ($E_p=E_W$) powerfully suggests that the holographic state is an optimal purification\footnote{Here we mean optimal in the sense of satisfying the minimization constraint that defines the entanglement of purification.} of the density matrix of any geometric subregion of the boundary theory. In further work \cite{BaoHal}, it was shown that there exists a conditional generalization of the entanglement of purification (with a corresponding holographically dual object) that passes the same consistency checks as the $E_p=E_W$ conjecture. This conditional generalization in the holographic context suggests an interpretation of the portion of the entanglement wedge of a region $ABC$ excluding the entanglement wedge of a subregion $C$ as being related to a constrained purification of the density matrix $\rho_{AB}$ given that the purification must include $C$. Moreover, this conditional entanglement of purification can be shown to nontrivially upper bound the conditional mutual information in any quantum state.

In a similar spirit, one can ask if there exists a simple generalization of the entanglement of purification that would upper bound other multipartite entanglement combinations such as the tripartite information, shown to be positive in \cite{Hayden:2011ag} \footnote{In \cite{BaoHal} this was done using combinations of conditional entanglements of purifications, but that bounding was not tight.}. In this work, we will show that the answer to this question is yes. In fact, we find generalizations of the entanglement of purification that upper bound both the tripartite information and the cyclic combinations shown to be positive holographically in \cite{EntCone}. Indeed, we prove that these upper bounds hold in any quantum system, regardless of the existence of a holographic dual.

After the first version of this paper appeared, a definition of multipartite entanglement of purification differing from ours only by a factor was proposed in \cite{Umemoto}, where a conjectured holographic dual which differs from our multipartite generalization of $E_W$ was also proposed.

The organization of this paper is as follows. In section 2, we briefly review the definitions and properties of the standard entanglement of purification and entanglement wedge cross-section. In section 3, we review the definition of the conditional entanglement of purification, its conjectured holographic dual, and demonstrate a few new properties of both. In section 4, we define the multipartite entanglement of purification and multipartite entanglement wedge cross-section, and prove they they share several properties but, nonetheless, are not holographic duals. Finally, we conclude with some discussion in section 5.

\section{Preliminary Definitions}

Consider a bipartite quantum system $AB=A \cup B,$ where $A$ and $B$ are taken to be disjoint. In fact, unless of otherwise stated, any two regions will be taken to be disjoint throughout the paper. The {{entanglement of purification} $E_p(A:B)$ is defined by 

\begin{equation}
E_p(A:B)=
\min \{S(AA');\rho_{AA\rq{}BB\rq{}} \text{ is pure} \}
\end{equation}
where $S$ is the Von Neumann entropy. $E_p$ is known to satisfy the following inequalities \cite{Terhal,Bagchi}:
\begin{eqnarray}
\min (S_A, S_B)\geq E_p(A:B) \geq \frac{1}{2} I(A:B) \label{eq:Eulb}\\
E_p(A:BC) \geq E_p(A:B) \label{eq:Esub} \\
E_p(AB:C) \geq \frac{1}{2} \left(I(A:C) + I(B:C)\right) \label{eq:Emon},
\end{eqnarray}
where  $I(A:B)\equiv S(A)+S(B)-S(AB)$ is the mutual information between $A$ and $B$. 

The $E_p=E_W$ conjecture was motivated by the proofs in \cite{TakUme} that the above inequalities are also all satisfied by a holographic object, the entanglement wedge cross-section, $E_W,$ defined by: 

\begin{equation}
E_{W}(A:B) =
\min \{\Ar(\Gamma); \Gamma \subset r_{AB} \text{ splits } r_{AB} \text{ into two regions homologous respectively to }A \text{ and } B\},
\end{equation}
where $r_{AB}$ is the restriction of the entanglement wedge\cite{Headrick:2014cta} of $AB$ to some time-symmetric slice. This restriction will be left implicit in what follows. Also implicit in this definition, and any holographic statement, is the existence of a (semi)-classical bulk geometry. The concepts of $E_p$ and $E_W$ are illustrated in Figure (\ref{fig:EwEp}).

\begin{figure}[h]
    \centering
    \includegraphics[width=0.7\textwidth]{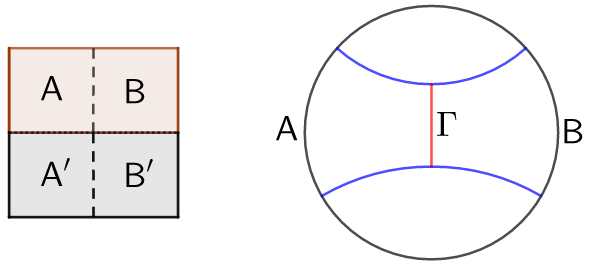}
    \caption{To the left, $A'B'$ purifies $AB.$ For a choice of $A'$ and $B'$ over all such purifying systems that minimizes the entanglement across the dashed partition we have $E_p(A:B)=S(AA').$ To the right,  $\Gamma$ is the minimal surface the separates the entanglement wedge cross-section of $AB$ into a region homologous to $A$ and a region homologous to $B.$ Its area is $E_W[A:B].$ [Figure adapted with permission from figure in \cite{BaoHal}].}
    \label{fig:EwEp}
\end{figure}

\section{The Conditional Entanglement of Purification}

In this section, we interpret quantities previously defined in \cite{BaoHal} as conditional entanglement of purification\footnote{In \cite{BaoHal}, what we now write as $E_p(A:B|C)$ was denoted $E_p(AC:BC),$ and likewise for $E_W.$ This new notation is to standardize with the new conditional interpretation. While this article was in preparation, reference \cite{Espindola} appeared using a notation suggestive of the conditional interpretation.} and conditional entanglement wedge cross-section. We also derive some new properties. 

When conditioned on subsystem $C,$ we get the {\bf conditional entanglement of purification} defined by 

\begin{equation}
E_p(A_1:A_2 | C)=\min_{A_1'A_2'C^{(1)}} \left\{ S(A_1A_1' C^{(1)}), \text{s.t. } \rho_{A_1 A_2 A_1' A_2' C} \text{ is pure and }C^{(1)}\subset C\right\},
\end{equation}
and the {\bf conditional entanglement wedge cross-section} by
\begin{equation}
E_W(A_1:A_2|C)=\min_{ \Gamma \in r(A_1 A_2 C)\backslash r(C)} \left\{\Ar(\Gamma), \text{s.t. } \Gamma \text{ splits } r(A_1 A_2C)\backslash r(C) \text{ accordingly} \right\}.
\end{equation}

In the spirit of the new conditional interpretation, one can also prove conditional analogs of  (Eqs. (\ref{eq:Eulb})--(\ref{eq:Emon})); the first two of these were proven in \cite{BaoHal}, but the third was missed there due to \cite{BaoHal} not referencing the conditional interpretation. Nevertheless, it is straightforward to show using the techniques of \cite{BaoHal} that its conditional generalization holds, i.e.
\begin{equation}
E(A:BC|D) \geq \frac{1}{2}I(A:B|D)+\frac{1}{2}I(A:C|D),
\end{equation}
where $E$ here can stand for either $E_p$ or$E_W$.

Furthermore, the following inequality may be dubbed the super-Bayesian property\footnote{Note this is also a generalization of the ``strong super-additivity'' inequality from \cite{TakUme}.}:
\begin{equation}
E_W(A_1 B_1: A_2B_2|C) \geq E_W(A_1:A_2|BC) + E_W(B_1:B_2|C), \label{eq:sBay}
\end{equation}
where $B=B_1 \cup B_2.$ The name is due to the resemblance with the Bayesian property of probabilities:
\begin{equation}
\ln p(AB|C) = \ln p(A|BC)+\ln p(B|C).
\end{equation}

It is easy to see that $E_W$ is super-Bayesian, i.e., it satisfies  Eq.~(\ref{eq:sBay}). This follows from the fact that the minimal surface that splits $A_1 B_1$ from $A_2 B_2$ in $r(ABC)\backslash r(C)$ can be broken into a piece that splits $A_1$ from $A_2$ in $r(A)\backslash r(BC)$ and one that splits $B_1$ from $B_2$ in $r(BC)\backslash r(C),$ and that a constrained optimizations is at most as optimal as a strictly less constrained optimization. See figure \ref{fig:sBay}.

\begin{figure}[h]  \centering
\includegraphics[width=0.3 \textwidth]{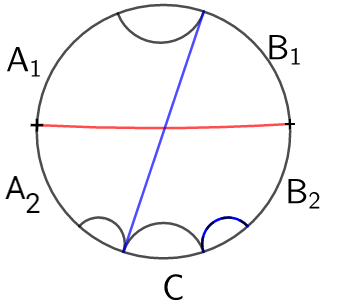} 
\caption{The red line is the minimal surface splitting $A_1B_1$ from $A_1B_2$ in $r(ABC)\backslash r(c),$ which is the region bounded by black lines. The RT surface of $BC$ is displayed in blue. It clearly splits the red line into two, one that splits (not necessarily minimally) $A_1$ from $A_2$ in $r(ABC)\backslash r(BC)$ and one that splits (not necessarily minimally) $B_1$ from $B_2$ in $r(BC)\backslash r(C)$.} \label{fig:sBay}
\end{figure}

On the other hand, $E_p$ is not super-Bayesian for arbitrary quantum states. For instance, one can construct a counterexample by choosing the regions to be subsystems of of GHZ states \cite{greenberger1989going}. If the $E_p=E_W$ conjecture is correct, however, $E_p$ must be super-Bayesian for any holographic state dual to a classical bulk geometry. This therefore allows for the super-Bayesian property to be used in an analogous way to the holographic entanglement entropy inequalities, i.e., as a discriminator of which states can be dual to (semi)-classical bulk geometries.

\section{The Multipartite Entanglement of Purification}

In this section, we define the {multipartite entanglement of purification} $E_p(A_1:A_2:\cdots :A_n)$ and the {multipartite entanglement wedge cross-section} $E_W(A_1:A_2:\cdots :A_n).$  Both of these quantities reduce to the bipartite objects when $n=2,$ and obey inequalities which reduce to some of those that motivated the formulation of the $E_p=E_W$ conjecture (Eqs. (\ref{eq:Eulb})--(\ref{eq:Emon})). This initially led the authors of this paper to conjecture that they were holographic duals. However, in section \ref{sec:EWproofs} we show this not to be the case.

The {\bf multipartite entanglement of purification} is defined by
\begin{equation}
E_p(A_1:A_2:\cdots :A_n)=\min_{A'} \left\{\frac{1}{n} \sum_{i=1}^n S(A_i A'_i), \text{such that } \rho_{A A'} \text{ is pure}\right\}, \label{eq:MEp}
\end{equation}
where $A = \cup_i A_i$ and $A'=\cup_i A_i'.$ (See Fig. \ref{fig:defs} for an example). It is clear that when $n=2,$ we get back the usual $E_p(A_1:A_2).$ 

Likewise, we define the {\bf multipartite entanglement wedge cross-section} $E_W(A_1:A_2:\cdots :A_n)$ as
\begin{equation}
E_W(A_1:A_2:\cdots :A_n)=\min_{ \Gamma \in r(A)} \left\{\frac{2}{n} \Ar(\Gamma), \text{such that } \Gamma \text{ splits } r(A) \text{ into } n  \text{ regions homologous to each }A_i \right\}. \label{eq:EWdef}
\end{equation}
See Fig. \ref{fig:defs} for an example with three regions.

\begin{figure}[h]  \centering
\includegraphics[width=0.7 \textwidth]{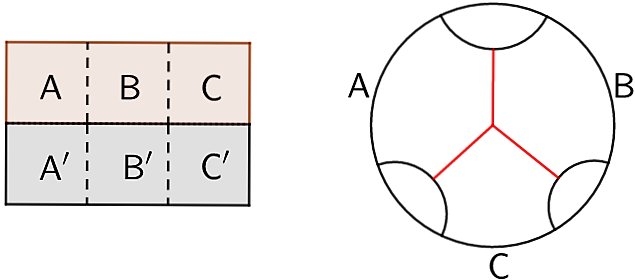} 
\caption{On the left,  $A'B'C'$ purifies $ABC$ while minimizing $\frac{1}{3}\left(S(AA')+S(BB')+S(CC')\right).$ This minimal value is  $E_p(A:B:C).$ On the right, the red surface is the minimal surface separating $r(ABC)$ into three regions, one homologous to $A,$ one to $B$ and one to $C.$  Its area is $\frac{3}{2}$ of $E_W(A:B:C).$} \label{fig:defs}
\end{figure}

We now study whether the inequalities (\ref{eq:Eulb})--(\ref{eq:Emon}) obeyed by $E_p$ and $E_W$ in the bipartite ($n=2$) case can be extended to the general multipartite case. Equations (\ref{eq:Eulb}) and (\ref{eq:Esub}) can be generalized in the multipartite case to

\begin{equation}
\frac{2}{n} \left(\sum_{i=1}^n S(A_i) - \max_i S(A_i) \right)  \geq E(A_1:\dots:A_n) \geq \frac{1}{n} C_k(A_1:\dots:A_n), \text{  and} \label{eq:ulb}
\end{equation}

\begin{equation}
E(A_1:A_2:\dots :A_i B:A_{i+1}:\dots:A_n) \geq E(A_1:A_2:\dots : A_i :A_{i+1}:\dots:A_n), \label{eq:monot}
\end{equation}
where $n=2k+1$ and $C_k$ is the cyclic information \cite{BaoHal, EntCone}defined as 
\begin{equation}
C_k(A_1: \dots :A_n) \equiv \sum_{i=1}^n \left( S(A_iA_{i+1}\dots A_{i+k})-S(A_i)\right)-S(A_1 \dots A_n),
\end{equation}
where indices are interpreted mod $n.$

We will show these to hold for both $E=E_p$ and $E=E_W.$ When $k=1,$ the cyclic information reduces to the tripartite information, $I_3,$ and equation (\ref{eq:ulb}) provides a novel way of upper bounding it. One might also try to generalize Eq.(\ref{eq:Emon}) to the multipartite case as follows:

\begin{equation}
E(A_1:\dots :A_i B:\dots:A_n) \geq \frac{1}{n} \left( C_k(A_1:\dots : A_i :\dots:A_n) + C_k(A_1:\dots B :\dots:A_n)  \right). \label{eq:EtoC}
\end{equation}

However, as of now, we have not been able to prove Eq.~(\ref{eq:EtoC}) for arbitrary $n.$ We will, nonetheless, prove it for $n=3,$ which gives us another inequality involving tripartite information:

\begin{equation}
E(A:BC:D) \geq \frac{1}{3} \left( I_3(A:B:D) + I_3(A:C:D)  \right). \label{eq:EtoI3}
\end{equation}

\subsection{Proof of inequalities for multipartite $E_p$}

To show the upper bound in Eq.~(\ref{eq:ulb}), consider without loss of generality that $S(A_n)=\max_i S(A_i),$ and let $A'=A_n',$ i.e., consider a purification with $A_i=\emptyset$ for $i\leq n-1.$ Then, we get

\begin{eqnarray}
E_p(A_1:\dots:A_n) \leq \frac{1}{n} \left(\sum_{i=1}^{n-1} S(A_i) + S(A_nA_n')\right) &=& \frac{1}{n} \left( \sum_{i=1}^{n-1} S(A_i) + S(A_1 A_2 \dots A_{n-1}) \right) \nonumber \\ &\leq& \frac{2}{n} \sum_{i=1}^{n-1} S(A_i),
\end{eqnarray} 
where we have used that $A A'$ is pure, and subadditivity. To show the lower bound in Eq.~(\ref{eq:ulb}), consider an optimal purification $A'=\cup_i A_i'.$ Then,

\begin{eqnarray}
& &n E_p(A_1:\dots:A_n) + \sum_i S(A_i \dots A_{i+k}) +S(A_1 \dots A_n) =  \sum_{i=1}^{n} S(A_i A_i')  + \sum_i S(A_i \dots A_{i+k-1}) +S(A_1 \dots A_n) \nonumber\\ 
&=&\sum_{i=1}^n \left[ S(A_i A_i') + S(A_{i+1}\dots A_{i+k}) \right] +S(A_1 \dots A_n) 
\geq \sum_{i=1}^n S(A_i' A_i  A_{i+1} \dots A_{i+k})  +S(A_1 \dots A_n) \nonumber \\
&\geq& \sum_{i=2}^n S( A_i' A_i A_{i+1} \dots A_{i+k}) + S(A_1 A_2 \dots A_N A_1') +S(A_1\dots A_{1+k}) \nonumber\\
&\geq& \sum_{i=1}^n S(A_i  A_{i+1} \dots A_{i+k}) + S(A A') = \sum_{i=1}^n S(A_i  A_{i+1} \dots A_{i+k})
\end{eqnarray} 
where indices are mod $n,$ and we have used strong subadditivity and the fact that $AA'$ is pure. Rearranging the terms gives the lower bound for $E$ in Eq.~(\ref{eq:ulb}).

Monotonicity, Eq. (\ref{eq:monot}), follows from the fact that the quantity on the right-hand side of the inequality is defined as the solution of a less constrained optimization problem. Since $B$ can be considered as part of $A_k',$ any purification of $AB$ is also a purification of $A.$  Interestingly, when $k=1,$ this gives a novel bound on the tripartite information $I_3=C_1. $

Let's now show that tripartite $E_p$ satisfies Eq.~(\ref{eq:EtoI3}). Let  $$E(A:BC:D)=\frac{1}{3} \left(S(AA') + S((BC)(BC)') + S(DD') \right)$$ for some purification. Then,

\begin{eqnarray}
&& 3 E(A:BC:D)+2S(A)+S(B)+ S(C)+2S(D)+S(ABD)+S(ACD) \nonumber\\
&=&S(AA')+S((BC)(BC)')+S(DD') +2S(A)+S(B)+ S(C)+2S(D)+S(ABD)+S(ACD)\nonumber \\
&\geq& S(AA')+S((BC)(BC)')+S(DD')+ S(AC)+S(AD)+S(BD)+S(ABD)+S(ACD)\nonumber \\
&\geq& S(A A')+S((BC)(BC)')+S(AB)+S(D')+S(AC)+S(AD)+S(BD)+S(ACD) \nonumber \\
&\geq& S(A') + S(D') + S((BC)(BC)') + S(AB)+S(AC)+ S(AD) + S(BD)+ S(CD)  \nonumber \\
&\geq& S(A'D')+S(ADA'D')+S(AB) +S(AC)+ S(AD)+ S(BD) + S(CD) \nonumber\\
&\geq&   S(AD)+ S(CD) + S(AB) + S(AD)+ S(BD)+S(AC), \nonumber \\ \label{eq:proof}
\end{eqnarray}
where we have used subadditivity, weak monotonicity, and that $\rho_{ABCDA'(BC)'D'}$ is pure. A rearrangement of the above inequality gives ~(\ref{eq:EtoI3}).

\subsection{Proof of inequalities for multipartite $E_W$} \label{sec:EWproofs}

The upper bound in Eq.~(\ref{eq:ulb}) can be established by noticing that the union of $(n-1)$ Ryu-Takayanagi (RT) surfaces splits $r(A)$ into the desired $n$ regions. Picking the $(n-1)$ RT surfaces with the smallest areas gives us this bound. For a pictorial proof of the lower bound, see Figure \ref{fig:EWlb}.

\begin{figure}[h]
    \centering
    \includegraphics[width=0.33\textwidth]{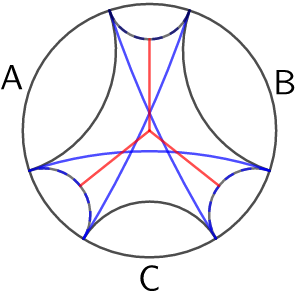}
    \caption{Visual proof of the lower bound in Eq.~(\ref{eq:ulb}) for $E_W$ shown for $k=1.$  The proof generalizes straightforwardly to arbitrary $k.$  Rearranging  Eq.~(\ref{eq:ulb}) so that terms on both sides of the inequality are all positive, we get $3 E_W(A:B:C) +S(A)+S(B)+S(C)+S(ABC) \geq S(AB)+S(AC)+S(BC).$ The black and the red lines correspond to the greater than (or equal) side of the inequality, with the red lines corresponding to the $3E_W$ term and being doubled (see definition of $E_W$ in Eq.~(\ref{eq:EWdef})). The blue lines correspond to the lesser than (or equal) side of the inequality. The dashed black-and-blue lines appear on both sides. By subadditivity, $S(A)+S(B)+S(C) \leq S(ABC),$ allowing us to replace the red lines with the dashed black-and-blue lines. Using these and each red segment twice, one can subtend each blue arc sub-optimally.}
    \label{fig:EWlb}
\end{figure}

Equation ~(\ref{eq:monot}) holds because, by entanglement wedge nesting \cite{Wall:2012uf, Akers:2016ugt}, we have that $r(AB) \supset r(A),$ and thus if a surface $\Gamma$ splits $r(AB)$ into $n$ regions homologous to each of $A_1, A_2, \dots, A_i B, \dots, A_n,$ then $\Gamma \cap r(A),$ which can have no greater area than $\Gamma,$ will split $r(A)$ into $n$ regions homologous to each of $A_1, A_2, \dots, A_i, \dots, A_n.$ 

Even though we do not know of a proof of the more general Eq.~(\ref{eq:EtoC}), we can prove Eq.~(\ref{eq:EtoI3}) holographically by following line by line Eq.~(\ref{eq:proof}).  This is not particularly illuminating, but it can be made rigorous using inclusion-exclusion techniques as in \cite{Hayden:2011ag}.  

Since multipartite $E_p,$ as defined by Eq.~(\ref{eq:MEp}), and multipartite $E_W,$ as defined by Eq.~(\ref{eq:EWdef}), obey the same set of inequalities, one may be led to believe these to be duals. Indeed, in a previous version of this paper, the authors conjectured this to be the case. However, a holographic tripartite pure state is a counterexample: if $\rho_{ABC}$ is pure, it follows that $E_p (A:B:C)=\frac{1}{3} \left( S(A)+S(B)+S(C)\right)$ \cite{Umemoto}, while the corresponding $E_W$ is generically larger than that (See Fig. \ref{fig:counter}). The geometric object conjectured in \cite{Umemoto} as dual to the multipartite entanglement of purification evades this counterexample, and it remains to see if it will endure future tests.

\begin{figure}[h]
    \centering
    \includegraphics[width=0.25\textwidth]{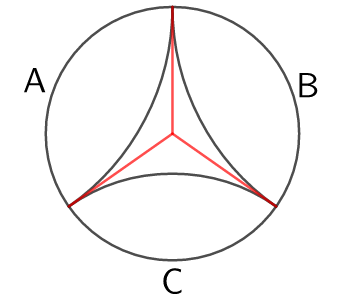}
    \caption{Counterexample showing that tripartite $E_p$ and $E_W$ are not dual to each other. Minimality of the RT surfaces imply that $E_W \geq \frac{1}{3} \left( S(A)+S(B)+S(C)\right),$ with the inequality being generically strict. Note, however, that a regulator is needed to makes sense of this statement since otherwise both sides of the inequality are divergent.}
    \label{fig:counter}
\end{figure}

\section{Discussion}
The multipartite and conditional entanglements of purification provide nontrivial upper bounds to known information theoretic quantities. In particular, in the context of the tripartite and cyclic informations, they give them new, Holevo-like \cite{Holevo1973bounds}, interpretations as the optimal multipartite entanglement of purification of some density matrix in any quantum system \footnote{For a recent holographic analysis of this, see for example \cite{Bao2017distinguishability}.}. Moreover, the fact that the conditional $E_p=E_W$ conjecture seems to produce nontrivial results in both the bulk and boundary increases the plausibility of the correctness of the original $E_p=E_W$ conjecture.

The super-Bayesian inequality in the context of the conditional entanglement of purification is another example of an inequality that is only true holographically (i.e. not for an arbitrary quantum state), much like the strong superadditivity in \cite{TakUme}. As such, it can be used as another discriminator for which quantum states are permitted to have holographic duals. 

It has recently been proposed by \cite{Hirai:2018jwy} that the entanglement of purification can be calculated in 2D CFTs. If this is successful, it would be an interesting future direction to see if that method can prove the $E_p=E_W$ conjecture or, beyond that, any further generalization. Furthermore, perhaps $E_W$ surfaces anchored to boundary-anchored HRT  surfaces \cite{Hubeny:2007xt} that probe behind the event horizon of black holes formed from collapse \cite{Hubeny:2002dg} have areas which can be calculated both holographically and directly in the boundary field theory, providing a nontrivial check of the smoothness of the region behind the black hole horizon in black holes formed by collapse.

Finally, the usage of the conditional and multipartite entanglements of purification to partition bulk minimal surfaces is of great use in assigning Hilbert space factors to different subregions of the bulk, and will be something that will be used to some effect in building tensor networks via entanglement distillation \cite{BaoWall}.

\section*{Acknowledgments}
We thank James Sully and Aron Wall for discussions.
This work is supported in part by the Berkeley Center for Theoretical Physics.
N.B. is supported by the National Science Foundation, under grant number 82248-13067-44-PHPXH.
 I.F.H. is supported by National Science Foundation under grant PHY-1521446.
\bibliographystyle{utcaps}
\bibliography{all}

\end{document}